\documentclass[]{spie}  

\usepackage{amsmath,amsfonts,amssymb}

\usepackage{float}
\usepackage{graphicx}
\usepackage{setspace}
\usepackage{tocloft}
\usepackage{lineno}
\usepackage[colorlinks=true, allcolors=blue]{hyperref}

\title{Single-shot focal plane wavefront sensing with the spatially-clipped self-coherent camera}

\author[a, b]{Joshua Liberman}
\author[c, b]{Sebastiaan Y. Haffert}
\author[b]{Jared R. Males}
\author[a,b]{Kevin Derby}
\author[b]{Ewan S. Douglas}

\affil[a]{James C. Wyant College of Optical Sciences, University of Arizona,
Meinel Building 1630 E. University Blvd., Tucson, AZ. 85721}
\affil[b]{Steward Observatory, University of Arizona, 933 N Cherry Ave, Tucson, AZ, USA 85719}
\affil[c]{Leiden University, Leiden Observatory, The Netherlands}

\authorinfo{Further author information: (Send correspondence to Joshua Liberman)\\Joshua Liberman: E-mail: jliberman@arizona.edu}

\begin{document} 
\maketitle

\begin{abstract}
The Habitable Worlds Observatory requires active speckle suppression to directly image Earth-like exoplanets. Focal plane wavefront sensing and control allows us to detect, and subsequently remove, time-varying speckles through measurements of the electric field. Two measurement-based wavefront sensing approaches are pairwise probing (PWP) and the self-coherent camera (SCC). However, the PWP technique is time-consuming, requiring at least 4 images and reducing the speed at which aberrations can be eliminated. In the SCC, a coronagraph mask diffracts light outside of the Lyot stop, where it is filtered with a pinhole. The filtered light creates a reference beam, interfering with speckles that leak through the coronagraph. The classic implementation of the SCC only works over small spectral bandwidths and needs significantly oversized optics which limits its implementation. We propose a new variant, the Spatially-Clipped Self-Coherent Camera (SCSCC). The SCSCC utilizes a pinhole placed closer to the Lyot Stop, reducing the overall beam footprint and boosting the sensor resolution. A knife-edge beam splitter downstream of the Lyot Stop splits the light into two channels: fringed and unfringed. This allows us to sense the wavefront with a single exposure. Time-varying aberrations are effectively frozen in place, making them easy to remove. We present monochromatic simulation results of the SCSCC in a sensing and control loop, demonstrating a normalized intensity of $\approx 4*10^{-10}$ in a 5-20 $\lambda / D$ dark hole. We find that wavefront control paired with the SCSCC achieves $\approx50$x deeper contrast than that achieved with PWP in a temporally evolving speckle field. Our results make the SCSCC a valuable wavefront sensor concept for the upcoming Habitable World Observatory mission.
\end{abstract}


\section{INTRODUCTION}
\label{sec:intro}  

The direct imaging technique for removing starlight and detecting a new planet is the most promising method for observing and characterizing Earth-like worlds. As such, NASA is developing the Habitable Worlds Observatory (HWO)--a space-based direct imaging telescope to identify habitable planets\cite{pueyo2019}. In direct imaging, starlight is suppressed using a coronagraph optical system: A focal plane mask diffracts the on-axis light outward where a Lyot stop then blocks the diffracted beam\cite{kenworthy_2025}. By removing starlight and preserving planet light, astronomers can search for atmospheric biosignatures assessing a planet's habitability. However, planet-to-star flux contrast for Earth-like worlds is $\sim 10^{-10}$, making the direct imaging technique difficult to implement. Furthermore, slow-varying optical aberrations, or quasi-static speckles, cause starlight to leak through a coronagraphic system. Speckles induced by optical aberrations are the limiting source of noise, preventing astronomers from reaching deeper contrasts\cite{currie2023}. To mitigate aberrations, NASA's HWO will require extremely stable optics, presenting engineering challenges and budget constraints\cite{coyle2023}.

Higher order wavefront sensing and control (HOWFSC) is an optimal strategy for removing quasi-static speckles and reaching deeper contrasts. In HOWFSC, a deformable mirror is used to sense and subsequently minimize the electric field in a given region of the science image. This process is commonly referred to as "digging a dark hole." In the past, model-based techniques such as pairwise probing + electric field conjugation (EFC) have been commonly used\cite{giveon2011, groff2015}, with the aforementioned method being baselined for use in the Roman Space Telescope's coronagraphic instrument\cite{cady2025}. However, model-based HOWFSC approaches are either too slow or rely on optical models that break down on-sky. Correcting aberrated wavefronts and achieving contrasts sufficient for detecting Earth-like planets instead requires measurement-based HOWFSC techniques.

Two alternative wavefront sensing and control approaches are implicit electric field conjugation (iEFC) and the self coherent camera (SCC) + EFC. The iEFC technique is time-consuming, requiring at least 4 images and reducing the speed at which aberrations can be eliminated\cite{desai2024}. The SCC is both a coronagraph and a wavefront sensor, measuring the wavefront with just a single image. When combined with EFC, the SCC can measure and control wavefronts at least 4x faster than iEFC.

In the conventional SCC, diffracted light is spatially filtered through a pinhole in the Lyot stop. The filtered light interferes with the quasi-static speckle field to form fringes in the detector plane. These fringes are then used to sense and remove aberrations\cite{baudoz2006}. While the conventional SCC has existed for 20 years, it has rarely been implemented outside of a lab setting due to its narrow spectral range\cite{coyle2021, galicher_2019}. As such, multiple variations on the classic SCC have been developed, utilizing a broader bandwidth by moving the pinhole closer to the Lyot pupil. This modification to the SCC's optical design is enabled by modulating the pinhole and computing a difference image between the modulated and unmodulated beams. The various modulated SCC designs include: the Fast Modulated SCC (temporal modulation)\cite{martinez2019}, the Polarization Encoded SCC (polarization modulation)\cite{bos2021}, and the Spectrally Modulated SCC (wavelength modulation)\cite{haffert2022}. However, these wavefront sensors are either limited by their modulation speed or require extensive calibrations in post-processing, limiting their efficacy on-sky.

Our SCC flavor--referred to as the Spatially-Clipped SCC (or SCSCC)--utilizes empirical measurements to sense aberrations in a single shot. Time-varying speckles are effectively frozen in place, making them easy to remove. The SCSCC design provides a $\approx3x$ improvement in bandwidth over the classical SCC by utilizing a pinhole placed $\approx3x$ closer to the Lyot stop\cite{bos2021}. Implementing the SCSCC on NASA's Habitable Worlds Observatory will enable high-speed, active control of wavefront aberrations. 

In Section~\ref{sect:theory}, we present the SCSCC design and wavefront estimation formalism, In Section~\ref{sect:methods}, we detail our simulation approach and our temporal wavefront error implementation, in Section~\ref{sect:analysis}, we analyze our results, and in Section~\ref{sect:Conclusion}, we discuss and conclude our work.

\section{Theory}
\label{sect:theory}

\begin{figure}[H]
\begin{center}
\begin{tabular}{c}
\includegraphics[height=8cm,]{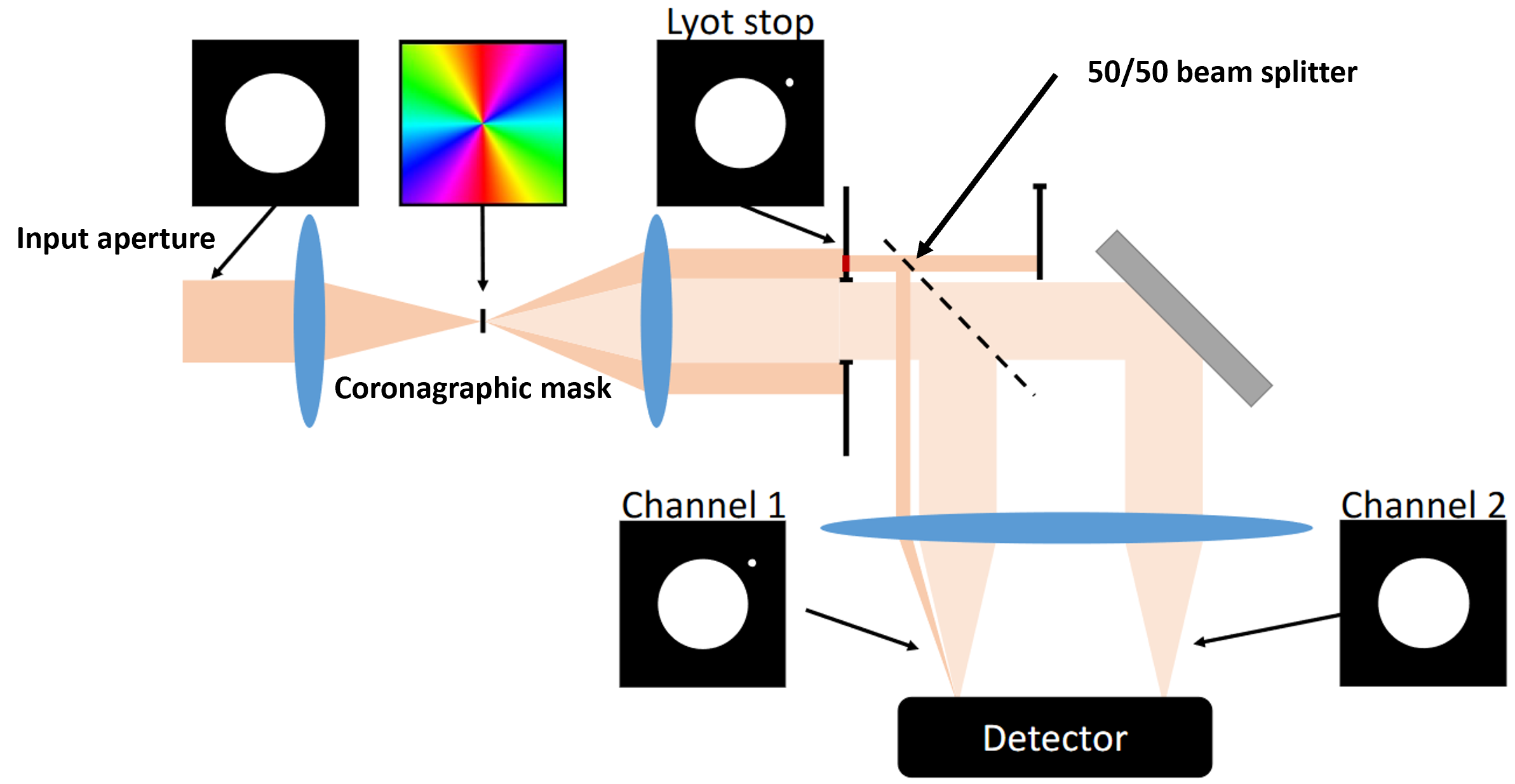}
\end{tabular}
\end{center}
\caption
{ \label{fig:scc-design}
An overview of the spatially-clipped SCC architecture. On-axis light focuses onto a coronagraph mask. The beam is then diffracted outwards and spatially filtered through a pinhole in the Lyot stop. The filtered light hits a beamsplitter, with one beam going to the detector while the other beam is spatially clipped by a knife edge stop. Light passing through the Lyot stop will also be split into two channels. One beam interferes with the spatially filtered light to produce fringes in the focal plane (channel 1), while the other, unfringed beam is used in the PSF subtraction (channel 2).}
\end{figure}

The SCSCC concept is based on a variation of the Spectrally Modulated SCC\cite{haffert2022}. A schematic of the SCSCC is shown in Fig.~\ref{fig:scc-design}. 

The mathematical formalism of the spatially clipped SCC may be expressed as follows. Let us define the complex speckle electric field after the Lyot stop as $\bar{E_{1}} = C\{E_{1}\}$ and the spatially filtered electric field (our reference) as $\bar{E_{2}} = C\{E_{2}\}$ where $C$ represents a coronagraph operator that propagates $E_{1}$ and $E_{2}$ through a coronagraphic optical system via a series of Fourier transforms. We may then express our encoded image channel formed by the interference between our speckle field ($\bar{E_{1}}$) and our reference field ($\bar{E_{2}}$) as

\begin{equation}
    I_{ch,1}=|\bar{E_{1}} + \bar{E_{2}}|^{2} = |\bar{E_{1}}|^{2} + |\bar{E_{2}}|^{2} + 2\Re\{\bar{E_{1}} \bar{E_{2}}^{\star}\}
\end{equation},
while our Lyot stop channel may be expressed as

\begin{equation}
    I_{ch,2}=|\bar{E_{1}}|^{2}
\end{equation}. Taking the difference between our two image channels, we obtain

\begin{equation}
    I_{ch,1} - I_{ch,2}= |\bar{E_{2}}|^{2} + 2\Re\{\bar{E_{1}} \bar{E_{2}}^{\star}\}
\end{equation}

where $\bar{E_{2}}^{\star}$ denotes the complex conjugate of $\bar{E_{2}}$. Now let us assume that $\bar{E_{2}}$, our field after the pinhole, is sufficiently small such that $\bar{E_{2}} << \bar{E_{1}}$. This assumption is reasonable, as our reference beam diameter is small and has a constant intensity.

To then minimize our final intensity and create a dark hole, we follow the methodology of Ref.~\citenum{thompson2022} and recover wavefront information from our fringes without taking a Fourier Transform. Re-writing our result in vector form, we obtain

\begin{equation}
\label{eqn:diff-img}
    \Delta I = 2 \begin{bmatrix}
\Re\{\bar{E_{2}}\} & \Im\{\bar{E_{2}}\} \\ 
\end{bmatrix} \begin{bmatrix}
 \Re\{\bar{E_{1}}\}\\ \Im \{\bar{E_{1}}\}
\end{bmatrix}
\end{equation}.

Now let us define a basis of Fourier modes, or sine waves, on the DM, such that $\Delta \bar{E_{M}}$ represents the change in the focal plane electric field through the SCC stop after applying a Fourier mode. The relationship between our excited mode states and our difference image from Eqn.~\ref{eqn:diff-img} may be linearly represented as

\begin{equation}
\label{eqn: lin-rep}
    \Delta I = 2 \begin{bmatrix}
\Re \{ \bar{E}_{M_{1}}\} & \Im \{ \bar{E}_{M_{1}}\} \\
\vdots & \vdots \\
 \Re \{ \bar{E}_{M_{n}}\}   & \Re \{ \bar{E}_{M_{n}}\}
\end{bmatrix} 
\begin{bmatrix}
 \Re\{\bar{E_{1}}\}\\ \Im \{\bar{E_{1}}\}
\end{bmatrix}
\end{equation}.

Alternatively, we may re-write Eqn.~\ref{eqn: lin-rep} as
\begin{equation}
    \Delta I = \textbf{B}a
\end{equation}
such that $\textbf{B}$ represents our interaction matrix of $n$ modes by $m$ pixels.

We may also view our interaction matrix \textbf{B} as a mapping between our modal coefficients ($a$) and our wavefront, following the methodology of Ref.~\citenum{haffert_implicit_2023}. Assuming linearity and an aberration-free system, we may decompose our matrix \textbf{B} into two transformation matrices \textbf{Z} and \textbf{C} where \textbf{Z} relates our modal coefficients to a phase and \textbf{C} relates our phase to a modulation in intensity as shown below: 

\begin{equation}
    \Delta I = \textbf{C}\textbf{Z}a
\end{equation}.

Computing the pseudo-inverse of \textbf{Z} and \textbf{C}, we obtain our mapping between the modal coefficients and modulated intensity: 
\begin{equation}
    a = \textbf{Z}^{-\dagger}\textbf{C}^{-\dagger}\Delta I
\end{equation}.

Our phase can then be expressed as a linear combination of our modal coefficients such that 
\begin{equation}
    \phi = \textbf{Z}a = \sum_n a_n Z_n
\end{equation}.

We may now compute a DM solution so as to minimize the DH intensity
\begin{equation}
    a = \arg\!\min_{a} |\Delta I+(\textbf{CZ})a|^{2} + \lambda |a|^{2}
\end{equation}
where $\lambda$ is a regularization parameter for penalizing solutions with large actuator stroke.

\section{Methods}
\label{sect:methods}
\begin{figure}[H]
\begin{center}
\begin{tabular}{c}
\includegraphics[height=6.5cm]{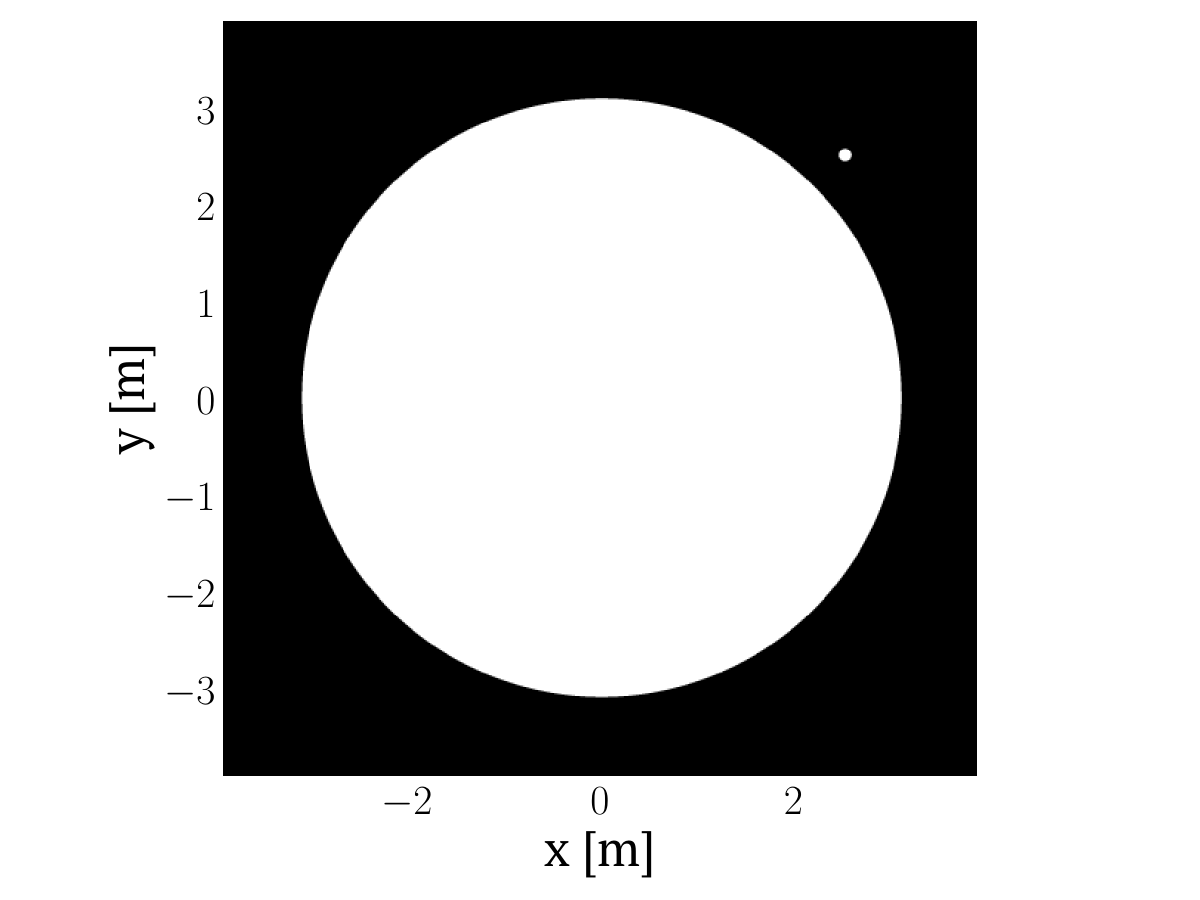}
\end{tabular}
\end{center}
\caption 
{ \label{fig:scc-stop}
The Lyot stop for the spatially clipped SCC. The pinhole is located at a radial distance of $54.5\% D_{\mathrm{EP}}$.} 
\end{figure}

To evaluate performance of the SCSCC, we developed a proof-of-concept physical optics simulation for a generic space-based coronagraphic instrument. We constructed our simulation with the High Contrast Imaging for Python (HCIPy) library\cite{por2018}. We generated a charge 4 scalar vortex coronagraph (SVC)\cite{ruane2019}, a circular telescope aperture, and a Lyot stop undersized by $95\%$ of the entrance pupil diameter ($D_\mathrm{EP}$). We perform wavefront control with a 52x52 actuator Boston Micromachines DM to simulate dark hole digging performance with a state-of-the-art DM available today.  Our SCC stop consists of a pinhole placed at a radial distance of $0.545 \cdot D_\mathrm{EP}$ from the center, with a pinhole diameter of $0.02\cdot D_\mathrm{EP}$. The simulated SCC stop is shown in Fig.~\ref{fig:scc-stop}. We generate spatially and temporally evolving speckles from spatiotemporal power spectral density (PSD) cubes. The spatiotemporal PSD is assumed to be separable, containing a spatial and a temporal component. We may express this as:
\begin{equation}
    S(f, k) = S_{t}(f)S_{s}(k).
\end{equation},
where $S_\mathrm{{t}}(f)$ denotes our temporal component and $S_{s}(k)$ denotes our spatial component. As such, each spatial frequency assumes the same temporal behavior. In reality, the temporal behavior varies as a function of spatial frequency\cite{males_mysterious_2021, poyneer_optimal_2006, macintosh_speckle_2005}. However, in this work, we assume the same temporal behavior for all spatial frequencies to reduce our explored parameter space.

We express our spatial component as
\begin{equation}
    S_{s}(f) = S_{0} * (f^{2} + f_{0}^{2})^{- \beta}
\end{equation}
where $f$ denotes our spatial frequency, and $S_\mathrm{0}$, $u_\mathrm{{0}}$, and $\beta$ are free parameters that describe the amplitude of wavefront error, the outer scale of the disturbance, and the distribution between high and low frequencies in the spatial PSD, respectively\cite{coulman1988}.

Our temporal component is expressed as 
\begin{equation}
    S_{t}(k) = (k^{2} + k_{0}^{2})^{- \alpha}
\end{equation},
where $k$ denotes the timestep, and $k_\mathrm{{0}}$ and $\alpha$ are free parameters that describe the correlation time of a disturbance and the roll-off of our power law, respectively.

\begin{figure}[H]
\begin{center}
\begin{tabular}{c}
\includegraphics[height=7.5cm]{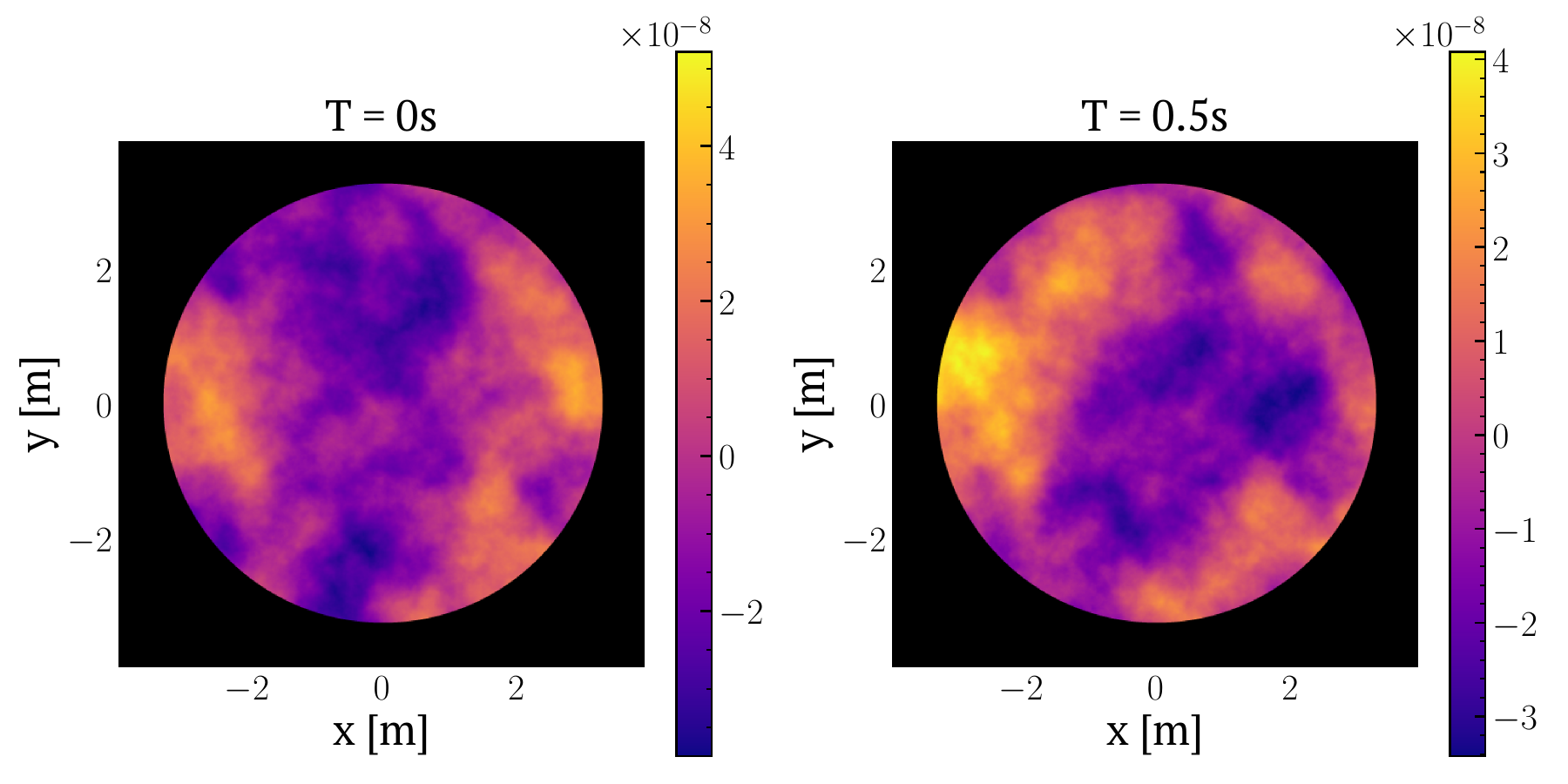}
\end{tabular}
\end{center}
\caption 
{ \label{fig:sa-temporal}
Phase screens depicting spatial and temporally varying surface aberrations across a pupil at time steps of 0 and 0.5 seconds. The color bar is given in units of meters.} 
\end{figure}

Speckles are defined over a spatiotemporal uniform grid with dimensions of 64 pix-by-64 pix-by-10 s. Optical phase screens for each time step are then generated from our PSDs with a random surface aberration distribution across the pupil. Sample phase screens for time steps of 0 and 0.5 s are shown below (Fig.~\ref{fig:sa-temporal}).

To construct the SCSCC interaction matrix, we use difference images between the PSFs propagated through the SCC stop, the Lyot stop with the pinhole blocked, and the pinhole alone. DM calibration modes are generated from a basis of Fourier modes with an amplitude of $0.01 \lambda$, where $\lambda = 750$nm. Computing the interaction matrix involves sequentially applying Fourier modes to the DM and using difference images from the modulated and unmodulated speckle fields to measure the resulting electric field for each mode in the focal plane. All images are normalized to unit power. We then pseudo-invert our matrix using Tikhonov regularization to generate a reconstruction matrix that maps the effect of each pixel in the detector plane to a corresponding Fourier mode in the pupil plane. We adopt a regularization strength of $2\cdot10^{-3}$ relative to the maximum singular value of our response matrix.

\begin{figure}[H]
\begin{center}
\begin{tabular}{c}
\includegraphics[height=6.5cm]{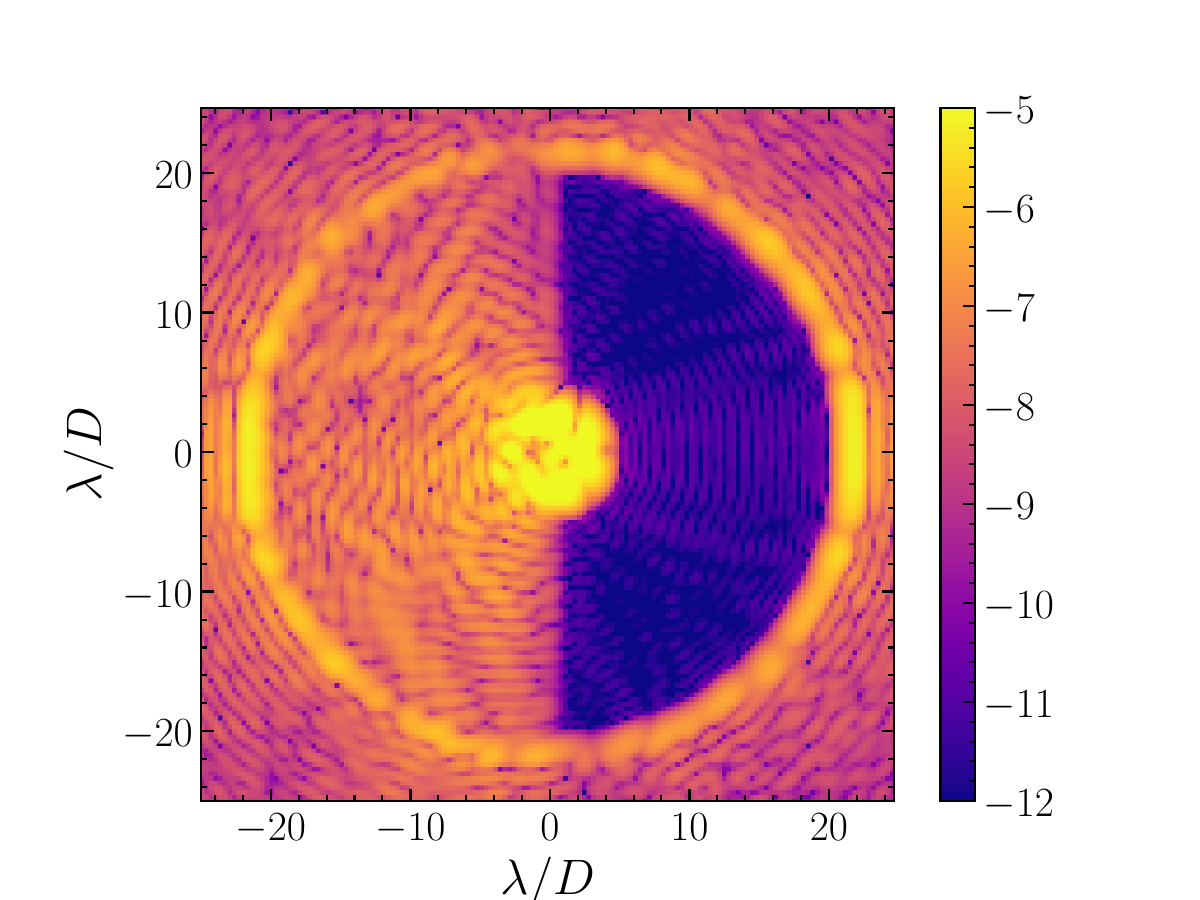}
\end{tabular}
\end{center}
\caption 
{ \label{fig:scc-psf}
A simulated 5-20$\lambda / D$ dark hole produced with the SCSCC + EFC. The colorbar denotes normalized intensity in log scale.} 
\end{figure}

A simulated stellar PSF after wavefront control with the SCSCC is shown in Fig.~\ref{fig:scc-psf}. In this example, we dig a D-shaped dark hole over 5-20$\lambda / D$ for a static speckle field. Note that the dark hole appears one-sided due to frequency folded speckles that fall within the dark hole region\cite{giveon_2006}.

\section{Analysis}
\label{sect:analysis}
\begin{figure}[H]
\begin{center}
\begin{tabular}{c}
\includegraphics[height=6.5cm]{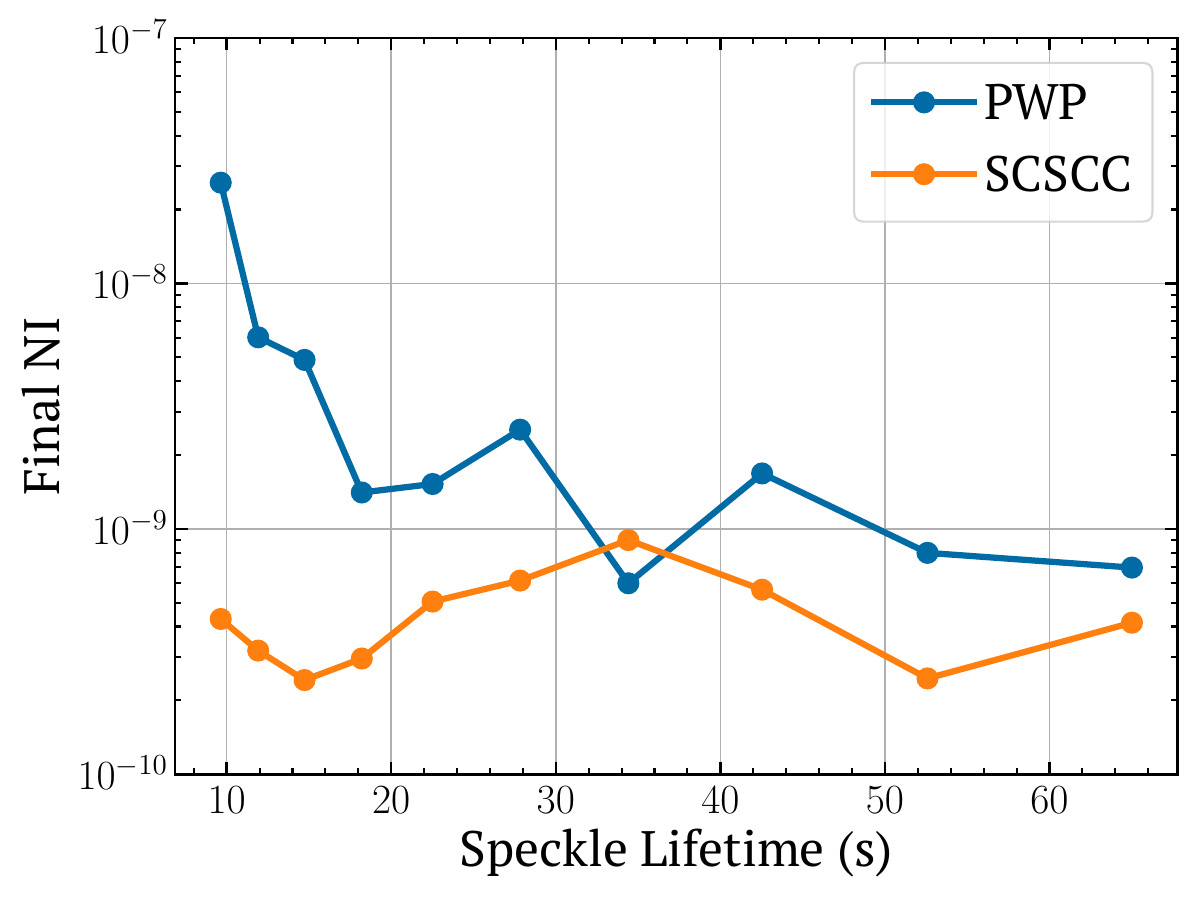}
\end{tabular}
\end{center}
\caption 
{ \label{fig:scc-pwp-ni}
A comparison of the SCC+EFC to PWP+EFC wavefront sensing and control methods as a function of speckle lifetime. The mean normalized intensity (NI) for the PWP+EFC method decreases for shorter lifetimes.} 
\end{figure}

We compared the SCSCC wavefront sensing performance to that of the PWP sensing strategy. In constructing our Jacobian from PWP measurements, we adopt the method used in \citenum{haffert2022}. We collect PWP measurements on a quasi-static speckle field, applying probes at a duty cycle of $100\%$. We adopt a probe amplitude of 0.01 $\lambda$. In creating our time-varying speckles for the SCC and PWP simulations, we choose fixed parameters of $\alpha = 4$, $\beta = 2$, and $u_{0} = 1$. To mimic aberrations that are consistent with space-like conditions, we use a peak-to-valley wavefront error amplitude of $0.075 \lambda$. Our results of the SCSCC and PWP comparison are shown in Fig.~\ref{fig:scc-pwp-ni}. 

For longer speckle lifetimes, both wavefront sensors sense and subsequently remove speckles at a mean NI of $\approx 10^{-10}$ with an effective integration time of 0.1 s per frame. This result is within the nominal contrast requirements for NASA's HWO mission. We observe that our SCSCC outperforms the PWP method by $\approx 1.7$ orders of magnitude at decreasing speckle lifetimes. This result is not surprising, as the SCSCC is capable of sensing aberrations from a single image as opposed to comparing modulated difference images in series.

\begin{figure}[H]
\begin{center}
\begin{tabular}{c}
\includegraphics[height=6.5cm]{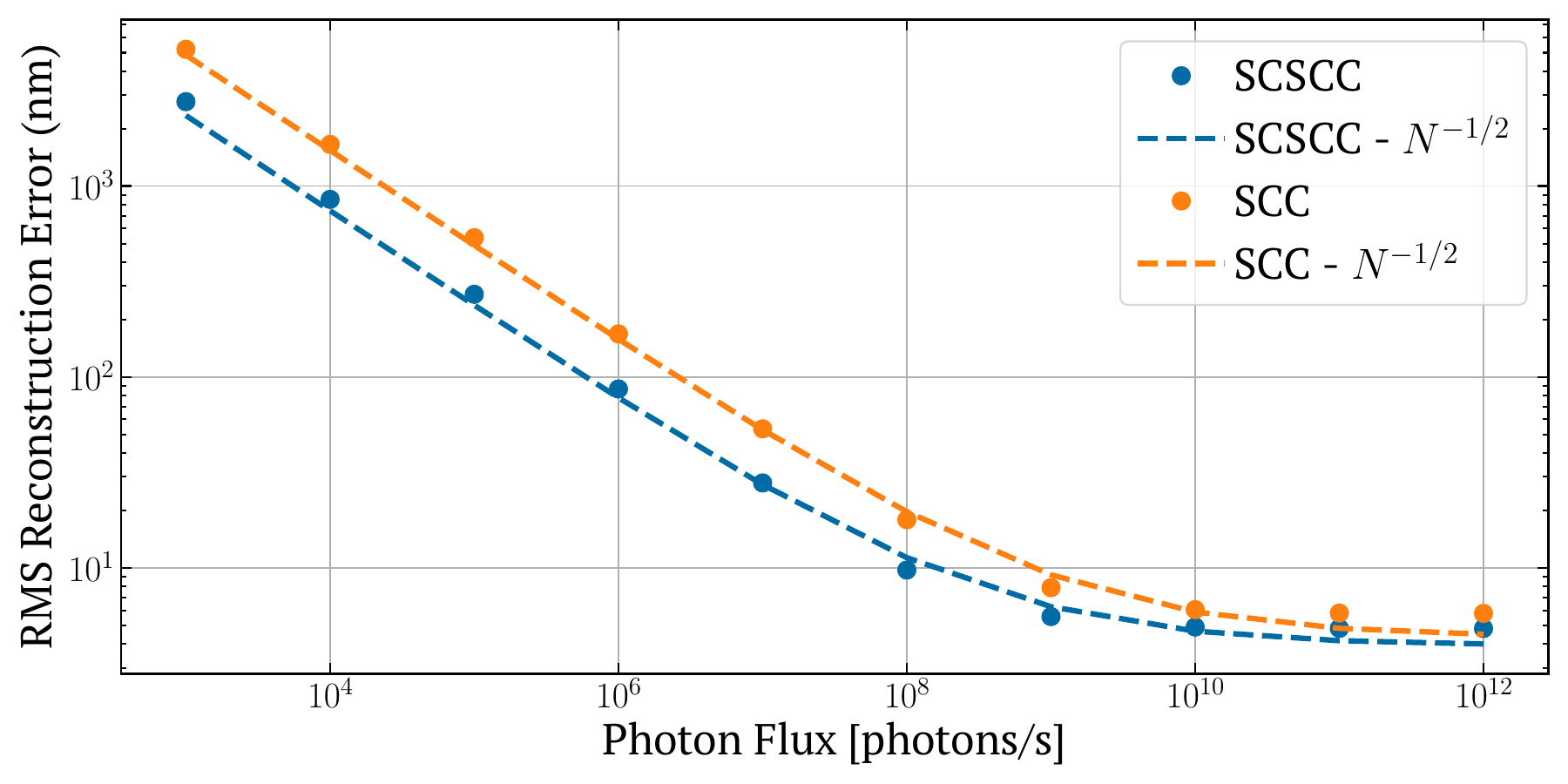}
\end{tabular}
\end{center}
\caption 
{ \label{fig:phot-noise}
Reconstructed wavefront RMS as a function of photon flux. The blue circles represent the SCSCC wavefront measurements while the orange circles represent the SCC measurement. The dashed lines represent $\propto \sqrt{N}$ curve fits with $N$ being the photon flux. Both wavefront sensors successfully recover the wavefront to within 5 nm RMS in the high photon flux regime.} 
\end{figure}

The SCSCC utilizes a pinhole placed much closer to the Lyot stop than the classical SCC. This produces a much higher flux through the reference channel of the SCSCC, thus increasing its sensitivity. Our SCC pinhole is placed at $1.545 \cdot D_\mathrm{EP}$ whereas our SCSCC pinhole is located at $0.545 \cdot D_\mathrm{EP}$. We follow the methodology of Ref.~\citenum{haffert2022} and numerically estimate the sensitivity gain provided by the SCSCC. We do so by injecting $\approx10$ nm RMS of static phase aberrations into our system with a power law of -2.5 and introducing photon noise. In both wavefront sensor calibrations, we adopt a regularization parameter of 0.08. We then compute the RMS of the difference between our reconstructed wavefront map and our input OPD to obtain the RMS reconstruction error. Fig.~\ref{fig:phot-noise} depicts RMS reconstruction error as a function of photon flux. 

We observe that both techniques reconstruct the wavefront to within 5 nm RMS, however, the accuracy can likely be improved by adjusting our regularization parameters and the probe amplitude. Following Ref.~\citenum{haffert2022}, we fit curves to our data points, with the expression $\frac{a}{\sqrt{\mathrm{N}}}+b$ where $a$ represents our photon efficiency and $b$ represents our asymptotic value. We compute the sensitivity gain by taking the ratio of $a$ between our SCC and our SCSCC curve fits. We observe a sensitivity gain of $\approx 2$x for the SCSCC. This trend differs from the results of Ref.~\citenum{haffert2022} where the authors observe a 5x sensitivity increase for the spectrally modulated SCC with a closer pinhole over the classical SCC. We believe this inconsistency stems from an imperfect tuning of our calibration.

\begin{figure}[H]
\begin{center}
\begin{tabular}{c}
\includegraphics[height=6.5cm]{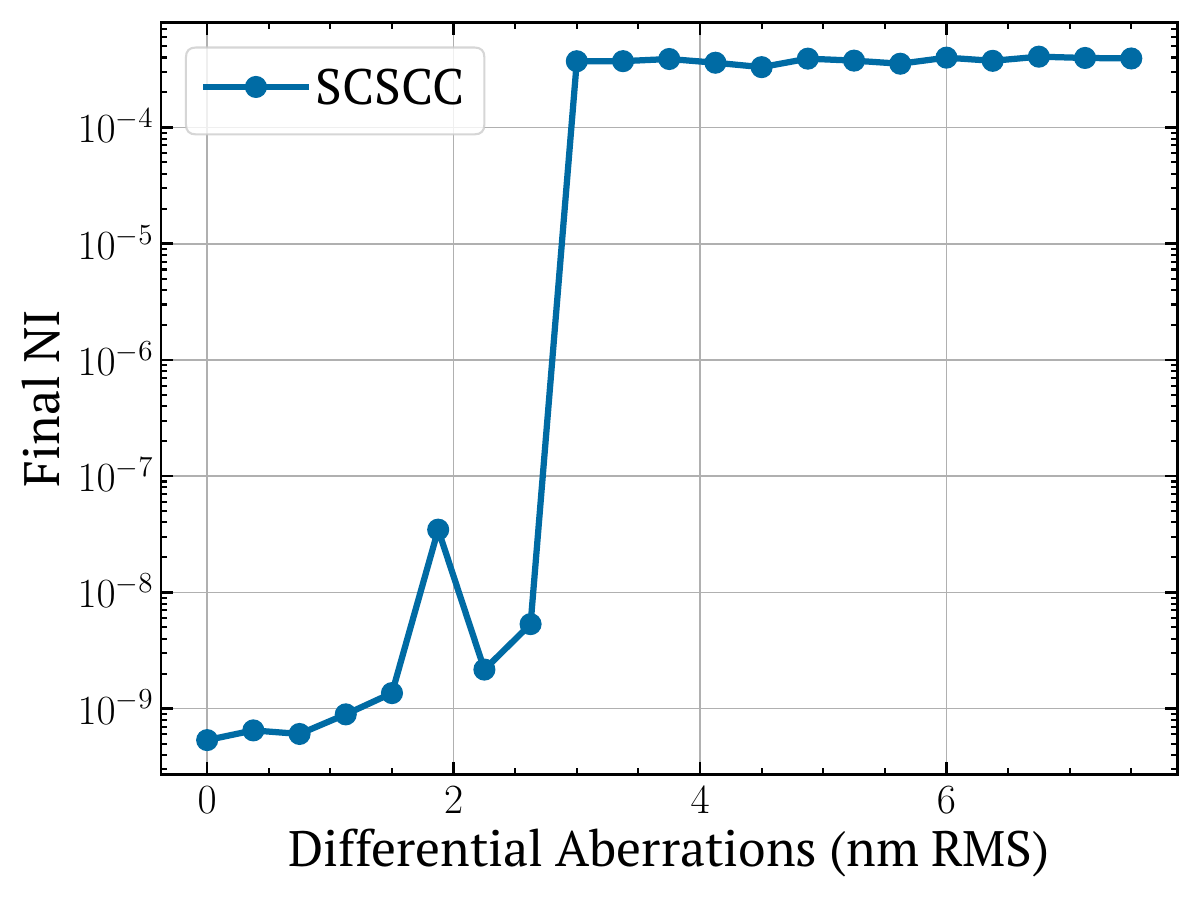}
\end{tabular}
\end{center}
\caption 
{ \label{fig:diff-aberrs}
Final NI as a function of differential amplitude aberrations for the SCSCC. The control loop begins diverging for differential aberrations greater than $\approx 2$ nm RMS.} 
\end{figure}

The SCSCC functions as a non-common-path wavefront sensor, with a 50/50 beam splitter dividing the light downstream of the Lyot stop between fringed and unfringed channels. As such, it is important to analyze the effects of differential aberrations between the two beam paths on our final contrast. In Fig.~\ref{fig:diff-aberrs}, we introduce identical static amplitude aberrations to our unfringed channel, such that they are present during and after calibration. We adopt a loop gain of 0.25 and a regularization of 0.03 with a quasi-static speckle field identical to that of Fig.~\ref{fig:scc-pwp-ni}.

We are able to achieve $\approx 10^{-10}$ contrasts when differential phase aberrations are $\leq 2$nm RMS. Beyond 2 nm RMS, we begin observing some instability in the control loop, with the solution fully diverging for differential aberrations $\geq 2.2$nm RMS. These results are roughly consistent with what we might expect, as a reconstruction RMS smaller than $\lambda/100$ necessitates that differential aberrations be smaller than $\lambda/200$\cite{haffert2022}. 


\section{Conclusion}
\label{sect:Conclusion}
In this work, we present the design and first-order simulation results of the SCSCC. We find that the SCSCC accurately senses quasi-static speckles that vary on short timescales ($\leq 15$s). Furthermore, we examine the SCSCC's sensitivity to differential aberrations between the fringed and unfringed beam paths, finding that the control algorithm diverges when differential aberrations exceed 2 nm RMS. We also show the SCSCCs improved wavefront reconstruction accuracy over the classical SCC in the reduced flux regime, noting a $\approx 2$x gain in sensitivity with the SCSCC. We can further improve the accuracy of our reconstruction with better tuning of our interaction matrix calibration.

Future work will involve developing the masks for the SCSCC and constructing the wavefront sensor on the Extreme Wavefront Control Laboratory's CACTI testbed\cite{schatz2022}. To this end, we have already procured a Hamamatsu ORCA-Quest2 qCMOS camera, enabling high-speed imaging with minimal read noise\cite{roth_qcmos_2024}. We also plan on implementing broadband capabilities for the SCSCC by utlizing SCC masks with multiple pinholes as demonstrated in Ref.~\citenum{desai2024}. We will then install the SCSCC on MagAO-X--an extreme adaptive optics system located on the Magellan-Clay 6.5m telescope. First light of the SCSCC on MagAO-X is planned for Fall 2026.

\acknowledgments   
I would like to thank Dr. Kian Milani and Dr. Niyati Desai for helpful discussions about the implementation of the spatially clipped SCC.

This work is supported by NASA APRA grant \#80NSSC24K0288. This research made use of community-developed core Python packages, including: HCIPy\cite{por2018}, Astropy\cite{robitaille2013}, Matplotlib\cite{hunter2007}, and the IPython Interactive Computing architecture\cite{perez2007}. Joshua Liberman is a member of UCWAZ Local 7065.
\bibliography{report} 

\begin{thebibliography}{10}

\bibitem{pueyo2019}
Pueyo, L., Stark, C., Juanola-Parramon, R., Zimmerman, N., Bolcar, M., Roberge, A., Arney, G., Ruane, G., Riggs, A. J.~E., Belikov, R., Sirbu, D., Redding, D., Soummer, R., Laginja, I., and Will, S., ``The luvoir extreme coronagraph for living planetary systems (eclips) i: searching and characterizing exoplanetary gems,''  3 (09 2019).

\bibitem{kenworthy_2025}
Kenworthy, M.~A. and Haffert, S.~Y., ``High-{Contrast} {Coronagraphy},'' (May 2025).
\newblock Publisher: Annual Reviews.

\bibitem{currie2023}
{Currie}, T., {Biller}, B., {Lagrange}, A., {Marois}, C., {Guyon}, O., {Nielsen}, E.~L., {Bonnefoy}, M., and {De Rosa}, R.~J., ``{Direct Imaging and Spectroscopy of Extrasolar Planets},'' in [{\em Protostars and Planets VII}{\nolinebreak\hspace{0.1em}]},  {Inutsuka}, S., {Aikawa}, Y., {Muto}, T., {Tomida}, K., and {Tamura}, M., eds., {\em Astronomical Society of the Pacific Conference Series} {\bf 534},  799 (July 2023).

\bibitem{coyle2023}
Coyle, L.~E., ``Enabling the {Habitable} {Worlds} {Observatory}: using systems engineering, integrated modeling and technology development to achieve ultra-stable optical systems,'' in [{\em Astronomical {Optics}: {Design}, {Manufacture}, and {Test} of {Space} and {Ground} {Systems} {IV}}{\nolinebreak\hspace{0.1em}]},  Hull, T.~B., Kim, D., and Hallibert, P., eds.,  {\bf PC12677},  PC1267702, SPIE (2023).
\newblock Backup Publisher: International Society for Optics and Photonics.

\bibitem{giveon2011}
Give'on, A., Kern, B.~D., and Shaklan, S., ``Pair-wise, deformable mirror, image plane-based diversity electric field estimation for high contrast coronagraphy,'' in [{\em Techniques and {Instrumentation} for {Detection} of {Exoplanets} {V}}{\nolinebreak\hspace{0.1em}]},   {\bf 8151},  376--385, SPIE (Sept. 2011).

\bibitem{groff2015}
Groff, T.~D., Riggs, A. J.~E., Kern, B., and Kasdin, N.~J., ``Methods and limitations of focal plane sensing, estimation, and control in high-contrast imaging,'' {\em Journal of Astronomical Telescopes, Instruments, and Systems}~{\bf 2},  011009 (Dec. 2015).
\newblock Publisher: SPIE.

\bibitem{cady2025}
Cady, E., Bowman, N., Greenbaum, A., Ingalls, J., Kern, B., Krist, J., Marx, D., Poberezhskiy, I., Riggs, A., Ruane, G., Seo, B.-J., Shi, F., and Zhou, H., ``High-order wavefront sensing and control for the {Roman} {Coronagraph} {Instrument} ({CGI}): architecture and measured performance,'' {\em Journal of Astronomical Telescopes, Instruments, and Systems}~{\bf 11} (Apr. 2025).

\bibitem{desai2024}
Desai, N., Potier, A., Redmond, S.~F., Ruane, G., Poon, P.~K., Riggs, A. J.~E., Noyes, M., and Prada, C.~M., ``Comparative laboratory study of electric field conjugation algorithms,'' {\em Journal of Astronomical Telescopes, Instruments, and Systems}~{\bf 10} (July 2024).
\newblock arXiv:2309.04920 [astro-ph].

\bibitem{baudoz2006}
{Baudoz}, P., {Boccaletti}, A., {Baudrand}, J., and {Rouan}, D., ``{The Self-Coherent Camera: a new tool for planet detection},'' in [{\em IAU Colloq. 200: Direct Imaging of Exoplanets: Science \& Techniques}{\nolinebreak\hspace{0.1em}]},  {Aime}, C. and {Vakili}, F., eds.,  553--558 (Jan. 2006).

\bibitem{coyle2021}
Coyle, L., Knight, J.~S., Hicks, B., Frater, E., Ho, J., Shugrue, J., and Jurczyk, S., ``Hardware demonstrations of component-level technologies for ultra-stable optical systems,'' in [{\em 2021 {IEEE} {Aerospace} {Conference} (50100)}{\nolinebreak\hspace{0.1em}]},   1--11 (Mar. 2021).
\newblock ISSN: 1095-323X.

\bibitem{galicher_2019}
Galicher, R., Baudoz, P., Delorme, J.-R., Mawet, D., Bottom, M., Wallace, J.~K., Serabyn, E., and Shelton, C., ``Minimization of non-common path aberrations at the palomar telescope using a self-coherent camera,'' ~{\bf 631},  A143.
\newblock Publisher: {EDP} Sciences.

\bibitem{martinez2019}
Martinez, P., ``Fast-modulation imaging with the self-coherent camera,'' {\em Astronomy \& Astrophysics}~{\bf 629},  L10 (Sept. 2019).
\newblock Publisher: EDP Sciences.

\bibitem{bos2021}
Bos, S.~P., ``The polarization-encoded self-coherent camera,'' {\em Astronomy \& Astrophysics}~{\bf 646},  A177 (Feb. 2021).
\newblock Publisher: EDP Sciences.

\bibitem{haffert2022}
Haffert, S.~Y., ``The spectrally modulated self-coherent camera ({SM}-{SCC}): {Increasing} throughput for focal-plane wavefront sensing,'' {\em Astronomy \& Astrophysics}~{\bf 659},  A51 (Mar. 2022).
\newblock Publisher: EDP Sciences.

\bibitem{thompson2022}
Thompson, W., Marois, C., Singh, G., Lardière, O., Gerard, B., Fu, Q., and Heidrich, W., ``Performance of the {Fast} {Atmospheric} {Self} {Coherent} camera at the {NEW}-{EARTH} lab and a simplified measurement algorithm,'' in [{\em Adaptive {Optics} {Systems} {VIII}}{\nolinebreak\hspace{0.1em}]},   {\bf 12185},  724--735, SPIE (Aug. 2022).

\bibitem{haffert_implicit_2023}
Haffert, S.~Y., Males, J.~R., Ahn, K., Van~Gorkom, K., Guyon, O., Close, L.~M., Long, J.~D., Hedglen, A.~D., Schatz, L., Kautz, M., Lumbres, J., Rodack, A., Knight, J.~M., and Miller, K., ``Implicit electric field {Conjugation}: {Data}-driven focal plane control,'' (Mar. 2023).
\newblock arXiv:2303.13719 [astro-ph] version: 1.

\bibitem{por2018}
{Por}, E.~H., {Haffert}, S.~Y., {Radhakrishnan}, V.~M., {Doelman}, D.~S., {van Kooten}, M., and {Bos}, S.~P., ``{High Contrast Imaging for Python (HCIPy): an open-source adaptive optics and coronagraph simulator},'' in [{\em Adaptive Optics Systems VI}{\nolinebreak\hspace{0.1em}]},  {Close}, L.~M., {Schreiber}, L., and {Schmidt}, D., eds., {\em Society of Photo-Optical Instrumentation Engineers (SPIE) Conference Series} {\bf 10703},  1070342 (July 2018).

\bibitem{ruane2019}
Ruane, G., Mawet, D., Riggs, A. J.~E., and Serabyn, E., ``Scalar vortex coronagraph mask design and predicted performance,'' in [{\em Techniques and {Instrumentation} for {Detection} of {Exoplanets} {IX}}{\nolinebreak\hspace{0.1em}]},   61 (Sept. 2019).
\newblock arXiv:1908.09786 [astro-ph].

\bibitem{males_mysterious_2021}
Males, J.~R., Fitzgerald, M.~P., Belikov, R., and Guyon, O., ``The mysterious lives of speckles. i. residual atmospheric speckle lifetimes in ground-based coronagraphs,'' ~{\bf 133}(1028),  104504.

\bibitem{poyneer_optimal_2006}
Poyneer, L.~A. and Macintosh, B.~A., ``Optimal fourier control performance and speckle behavior in high-contrast imaging with adaptive optics,'' ~{\bf 14}(17),  7499--7514.

\bibitem{macintosh_speckle_2005}
Macintosh, B., Poyneer, L., Sivaramakrishnan, A., and Marois, C., ``Speckle lifetimes in high-contrast adaptive optics,''  {\bf 5903},  170--177.
\newblock {ADS} Bibcode: 2005SPIE.5903..170M.

\bibitem{coulman1988}
Coulman, C.~E., Vernin, J., Coqueugniot, Y., and Caccia, J.~L., ``Outer scale of turbulence appropriate to modeling refractive-index structure profiles,'' {\em Applied Optics}~{\bf 27},  155--160 (Jan. 1988).
\newblock Publisher: Optica Publishing Group.

\bibitem{giveon_2006}
Give'on, A., Kasdin, N.~J., Vanderbei, R.~J., and Avitzour, Y., ``On representing and correcting wavefront errors in high-contrast imaging systems,'' ~{\bf 23}(5),  1063--1073.

\bibitem{schatz2022}
Schatz, L.~H., Codona, J., Long, J.~D., Males, J.~R., Pullen, W., Gorkom, K.~V., Chambouleyron, V., Close, L.~M., Correia, C., Fauvarque, O., Fusco, T., Guyon, O., Hart, M., Janin-Potiron, P., Johnson, R., Jovanovic, N., Lumbres, J.~R., Mateen, M., Sauvage, J.-F., and Neichel, B., ``Experimental demonstration of a three-sided pyramid wavefront sensor on the {CACTI} testbed,'' in [{\em Adaptive {Optics} {Systems} {VIII}}{\nolinebreak\hspace{0.1em}]},   {\bf 12185},  121852D, SPIE (Aug. 2022).

\bibitem{roth_qcmos_2024}
Roth, M.~M., ``{qCMOS} detectors and the case of hypothetical primordial black holes in the solar system, near earth objects, transients, and other high-cadence observations,'' ~{\bf 8}(11),  282 (2024).
\newblock Publisher: The American Astronomical Society.

\bibitem{robitaille2013}
{Astropy Collaboration}, {Robitaille}, T.~P., {Tollerud}, E.~J., {Greenfield}, P., {Droettboom}, M., {Bray}, E., {Aldcroft}, T., {Davis}, M., {Ginsburg}, A., {Price-Whelan}, A.~M., {Kerzendorf}, W.~E., {Conley}, A., {Crighton}, N., {Barbary}, K., {Muna}, D., {Ferguson}, H., {Grollier}, F., {Parikh}, M.~M., {Nair}, P.~H., {Unther}, H.~M., {Deil}, C., {Woillez}, J., {Conseil}, S., {Kramer}, R., {Turner}, J. E.~H., {Singer}, L., {Fox}, R., {Weaver}, B.~A., {Zabalza}, V., {Edwards}, Z.~I., {Azalee Bostroem}, K., {Burke}, D.~J., {Casey}, A.~R., {Crawford}, S.~M., {Dencheva}, N., {Ely}, J., {Jenness}, T., {Labrie}, K., {Lim}, P.~L., {Pierfederici}, F., {Pontzen}, A., {Ptak}, A., {Refsdal}, B., {Servillat}, M., and {Streicher}, O., ``{Astropy: A community Python package for astronomy},'' {\em Astronomy \& Astrophysics}~{\bf 558},  A33 (Oct. 2013).

\bibitem{hunter2007}
Hunter, J.~D., ``Matplotlib: A 2d graphics environment,'' {\em Computing in Science \& Engineering}~{\bf 9}(3),  90--95 (2007).

\bibitem{perez2007}
Perez, F. and Granger, B.~E., ``Ipython: A system for interactive scientific computing,'' {\em Computing in Science and Engineering}~{\bf 9}(3),  21--29 (2007).

\end{thebibliography}
\bibliographystyle{spiebib} 

\end{document}